\documentclass[12pt,a4paper]{article}
\usepackage{amsmath,amsthm,amsfonts}

\def\req#1{(\ref{#1})}
\newcommand{\D}{{\rm d}}

\date{ }
\author{I. Klich, A. Mann and M. Revzen}
\begin{document}
\title{A thick shell Casimir effect} \maketitle \vskip 2mm
\begin{center}

Department of Physics,
\\ Technion - Israel Institute of Technology, Haifa 32000 Israel
\footnote{e-mail:\\ klich@tx.techion.ac.il \\
ady@physics.technion.ac.il \\ revzen@physics.technion.ac.il}
\\

\end{center}
\begin{abstract}
We consider the Casimir energy of a thick dielectric-diamagnetic
shell under a uniform velocity light condition, as a function of
the radii and the permeabilities. We show that there is a range of
parameters in which the stress on the outer shell is inward, and a
range where the stress on the outer shell is outward. We examine
the possibility of obtaining an energetically stable configuration
of a thick shell made of a material with a fixed volume.
\end{abstract}
\bigskip


\section{Introduction}
It is well known that the fluctuations of electromagnetic fields
in vacuum or in material media depend on the boundary conditions
imposed on the fields. This dependence gives rise to forces which
are known as Casimir forces, acting on the boundaries. The best
known example for such forces is the attractive force experienced
by parallel conducting plates in vacuum \cite{Cas48}. Casimir
forces between similar, disjoint objects such as two conducting or
dielectric bodies are known in most cases to be attractive
\cite{Kenneth1} and are sometimes viewed as the macroscopic
consequence of Van der Waals and Casimir-Polder attraction between
molecules.

However, Boyer \cite{Bo68} showed that the zero point
electromagnetic pressure on a conducting shell is directed
outward\footnote{Assuming that zero point forces are in general
attractive, Casimir considered a semiclassical toy-model for the
electron in which the coulomb self-repulsion is balanced by a
Casimir type attraction. One of the consequences of Boyer's
calculation is that since the pressure on a conducting sphere is
outward it cannot balance the Coulomb repulsion in Casimir's
model.}. In this paper we address the question: can there exist a
compact ball for which the Casimir forces would not expand the
ball to infinity? We look for such behavior in a simple model.

In view of the dominance of the Casimir forces at the nanometer
scale, where the attractive force could lead to restrictive limits
on nanodevices, the study of repelling Casimir forces is of
increasing interest. Indeed, Boyer, following Casimir's
suggestion, studied inter plane Casimir force with one plate a
perfect conductor while the other is infinitely permeable. He
showed that, in this case the plates repel \cite{Bo74}. This
problem was reconsidered since in \cite{Alves,daSilva}.

In addition to the Casimir effect for a conducting spherical
shell, various cases of material balls where considered in the
literature. The case of a ball made of a dielectric material was
considered by many authors and exhibits strong dependence on
cutoff parameters
\cite{Milton80,Bordag99,Barton,Marachevsky00,Marachevsky01,Nesterenko00}.
The case of a dielectric-diamagnetic ball has also been
extensively studied, especially under the condition of
$\epsilon\mu=1$, which will be referred to as the uvl (uniform
velocity of light) condition, since its introduction by Brevik and
Kolbenstvedt \cite{Brevik82,Brevik98,Nesterenko99,Kl}. A medium
with the uvl property has in many cases cutoff-independent values
for the Casimir energy (see also \cite{Nesterenko99C,Milton99,IKA}
for the Casimir energy of a cylinder with uvl). A heuristic
argument for this statement goes as follows: the zero point energy
of the electromagnetic field in a uvl medium is a sum over the
eigenfrequencies of the system, and behaves for high frequencies
as $\sum_I \omega_I\sim \sum_I ck_I$, where the factor $c$ is
common to all the media and the geometric information on the
system enters via the allowed $k'$s. This expression in the limit
of high $k's$ behaves exactly as the vacuum energy with
$\epsilon\mu=c^{-2}$ and the sum becomes regularizable by
subtraction of the vacuum energy. In cases where this condition is
not fulfilled there are problems of UV divergent terms which are
proportional to the volume in which the velocity of light is
different from the velocity of light in the background.

We will use the uvl condition for our thick shell.

In all of the above mentioned cases, namely, the conducting
sphere, dielectric ball or a dielectric-diamagnetic ball with uvl,
the resultant pressure was found to be repulsive.

\begin{figure}\center{\input epsf \epsfxsize=2.0in \epsfbox{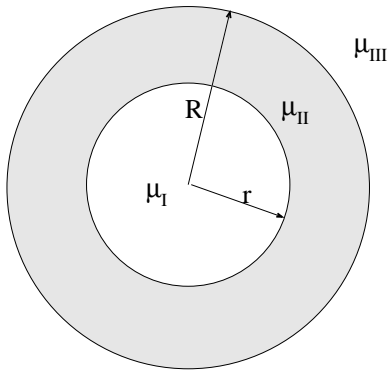}}
\caption{A thick dielectric-diamagnetic shell.} \label{f2}
\end{figure}

In order to obtain diverse behaviors we mix the case of disjoint
bodies (where there is usually attraction) and the dielectric ball
scenario as follows: We consider the case of a thick shell (fig.
\ref{f2}), with three permeability regimes (inner, middle and
outer) $\mu_I,\mu_{II},\mu_{III}$, under a uvl condition. In this
case there are two competing effects, i.e.: interaction between
the inner and outer boundaries, and the repulsive pressure
experienced by each boundary.

Using a formula derived in \cite{KlG} we show in sections
\ref{GREEN} and \ref{ENERGY} that for a dilute medium the energy
of the thick shell is $$E_C(r,R)=\frac{5}{128\pi R}\kappa_R^2 +
\kappa_r\kappa_R\frac{r^3(r^2-5R^2)}{4(r-R)^3(r+R)^3\pi}+\kappa_r^2\frac{5}{128\pi
r}$$ Where $r$ and $R$ are the inner and outer radii respectively
and the parameters $\kappa_r$ and $\kappa_R$ are defined in
section \ref{ENERGY}. From this expression one can easily obtain
the energy of a single ball by taking the limit of $R$ to
infinity, and also the energy of two parallel infinite dielectrics
by taking both of the radii to infinity while keeping a finite
distance $d=R-r$.

Next, in sections \ref{STRESS} and \ref{VOLUME} we investigate
this expression for the energy and show that there is a range of
parameters such that the force on the outer
shell is attractive. 


\section{Energy of the electromagnetic fluctuations}\label{GREEN}

In this section we briefly review the Green's function method for
calculation of Casimir energies (see \cite{LP} and \cite{KlG} for
details). To calculate the Casimir energy of a medium under the
uvl condition, we use the perturbative technique suggested in
\cite{KlG}. There, the Born series for the correlation function
$D_{ik}(\omega;{\bf r},{\bf r'})$ of the electromagnetic fields in
material medium is presented. The correlation function $D_{ik}$ is
defined by
\begin{equation}
D_{ik}(\omega;{\bf r},{\bf r'})=\int_0^{\infty}e^{i\omega t}
D^R_{ik}(t;{\bf r},{\bf r'})\D t
\end{equation}
where
$$D^R_{ik}(t_2-t_1;{\bf r},{\bf r'})=\Big{\{}\begin{array}{ll}
i<A_i({\bf r},t_1)A_k({\bf r'},t_2)-A_k({\bf r'},t_2)A_i({\bf
r},t_1)> \,\,\,& t_1<t_2 \\ 0  & {\rm otherwise}
\end{array}$$
is the retarded Green's function.  Throughout we use the gauge
$A_0=0$, so that the indices $i,k$ range $1,2,3$ and $D$ is a
$3\times 3$ matrix.

The correlation function $D$ is known \cite{LP} to be the Green's
function for the equation (in units where $\hbar=c=1$)
\begin{equation}\label{MainEquation}
([{\bf \nabla}]\mu^{-1}[{\bf
\nabla}]-\omega^2\epsilon)D=-{\bf{\mathbb I }}
\end{equation}
where $[a]$ stands for a matrix whose elements are
$[a]_{ik}=\epsilon_{ijk}a_j$ and ${\bf{\mathbb I }}$ is the
identity operator, which, in coordinate space is just the $3\times
3$ identity matrix times a delta function. In eq.
\req{MainEquation}, $\epsilon$ is the permittivity and $\mu$ the
permeability of the medium. In the following we assume that
$\mu({\bf r})$ and $\epsilon({\bf r})$ are scalar functions (of
course, in the general case both $\mu$ and $\epsilon$ are
tensors). This equation can be also written in the form:
\begin{equation}\label{eqDs}
([{\bf \nabla}]^2-[{\bf \nabla}\log\mu][{\bf
\nabla}]-\omega^2\epsilon\mu)D=-\mu{\bf{\mathbb I}}
\end{equation}

The correlation function of electromagnetic fields in vacuum,
where $\epsilon=\mu\equiv 1$, will be denoted $D_0$. It is the
inverse of the operator $([{\bf \nabla}]^2-\omega^2)$ and is well
known. It is given by the formula:
\begin{equation}\label{D00}
D_0(\omega;r,r')=-\left({\mathbb I }+{1\over\omega^2}{\bf
\nabla}\otimes{\bf \nabla}\right){1\over 4\pi|{\bf r}-{\bf
r}'|}\exp^{(i\omega |{\bf r}-{\bf r}'|)},
\end{equation}
We now wish to use the known $D_0$ to express $D$ via a Born
series. Define:
\begin{equation}
{\mathcal P}=\omega^2{\bf{\mathbb I }}(1-\mu\epsilon)
\end{equation}
and
\begin{equation}
{\mathcal Q}=-[{\bf \nabla}\log(\mu)][{\bf \nabla}]
\end{equation}
It follows from \req{eqDs} that as an operator $D$ satisfies:
\begin{equation}
D=((I- ({\mathcal P}+{\mathcal Q})
D_0){D_0}^{-1})^{-1}\mu=D_0(I-({\mathcal P}+{\mathcal Q})
D_0)^{-1}\mu,
\end{equation}
Thus $D$ is given by the following formal series:
\begin{equation}\label{expansion}
D=D_0\mu+D_0({\mathcal P}+{\mathcal Q})D_0\mu+D_0({\mathcal
P}+{\mathcal Q})D_0({\mathcal P}+{\mathcal Q})D_0\mu+....
\end{equation}

We now impose the "uniform velocity of light" condition in the
medium by setting $\epsilon\mu\equiv {\bf\rm I}$. This eliminates
the ${\mathcal P}$ terms and we are left with the expansion:
\begin{equation}\label{epansion}
D=D_0 \mu+D_0{\mathcal Q}D_0 \mu+D_0{\mathcal Q}D_0{\mathcal Q}D_0
\mu+....
\end{equation}

The correlation function $D$ can be used to calculate various
properties of the electromagnetic field. We use it to calculate
the energy density of the field by using the relations:
$$(E_iE_k)_{\omega}=\omega^2(A_iA_k)_{\omega} $$  and
$$(B_iB_k)_{\omega}=\overrightarrow{{\bf \nabla}}\times
(A_iA_k)_{\omega}\times\overleftarrow{{\bf \nabla}},$$ Where the
round brackets stand for the Fourier transform with respect to
time of the correlation function ${1\over 2}<A_i({\bf
r},t)A_k({\bf r'},0)+A_k({\bf r'},0)A_i({\bf r},t)>.$

At a temperature $T={1\over \beta}$ this correlation function of
the fields $A_i$ is related to the retarded Green's function $D$
by the fluctuation dissipation theorem \cite{LP}:
\begin{equation}
(A_i({\bf r})A_j({\bf r}'))_{\omega}=-i4\pi\coth({\beta\omega\over
2})( D_{ij}({\bf r},{\bf r}')-\overline{D_{ji}}({\bf r}',{\bf r}))
\end{equation}
Thus, at zero temperature, the energy density of the
electromagnetic field is:
\begin{eqnarray}\label{rhoTlim}
& \rho({\bf r},\omega)={1\over 4\pi}\Big(\epsilon({\bf r}) (E({\bf
r})^2)_{\omega}+{1\over \mu({\bf r})}(B({\bf
r})^2)_{\omega}\Big){\D\omega\over 2\pi}=\\ \nonumber & {1\over
4\pi}\lim_{{\bf r}'\rightarrow {\bf r}}{\bf\rm Im}\, {\rm Tr}
\Big(\omega^2\epsilon({\bf r})D({\bf r},{\bf r}')+{1\over\mu({\bf
r})}\overrightarrow{{\bf \nabla}}\times D
\times\overleftarrow{{\bf \nabla}}\Big)\D\omega,
\end{eqnarray}
where we have chosen to neglect the dependence of the permeability
$\mu$ and permittivity $\epsilon$ on the frequency $\omega$.
Inserting the expansion \req{epansion} in \req{rhoTlim} one can
obtain a series for the energy density of fluctuations of the
electromagnetic field. It was shown in \cite{KlG} that the first
contribution to the Casimir energy density (i.e., energy of the
electromagnetic field in the presence of external conditions minus
the energy density of the electromagnetic field without them)
comes from the term $D_0{\mathcal Q}D_0{\mathcal Q}D_0 \mu$ in
\req{epansion} \footnote{To see this, note that the $D_0$ term is
just the contribution of a homogeneous vacuum which we subtract.
The terms of the form $D_0{\mathcal Q}D_0{\mathcal Q}\mu$ can be
shown to cancel between the electric and magnetic correlation
functions.}. For a dilute medium this term gives the dominant
contribution to the Casimir energy.


\section{Casimir energy of a thick shell}\label{ENERGY}
The Casimir energy per $\D\omega$ is obtained via eq.
\req{rhoTlim} by the integration of $\rho({\bf r},\omega)-({\rm
contribution\,of}\,D_0)$ over space. This yields the following
general formula for the density of Casimir energy of a medium with
a radially symmetric permeability $\mu$ \cite{KlG} assuming the
uvl property, and diluteness:
\begin{equation}\label{maindensity}
{\rho_T}^{(2)}(\omega)= -4\pi\omega^2\coth({\beta\omega\over 2})
{\rm Im} \int_{0}^{\infty}{\rm d}s I(s)(\log\mu(s))'
\end{equation}
Where:
\begin{equation}
I(s)=\int_{\cal T}  {\rm dudv} {(u^4-(v^2-s^2)^2)\over 2u}
g_2(u)(\log\mu(v))'
\end{equation}
and the integration domain ${\cal T}$ is such that $u$, $v$ and
$s$ can form a triangle. The function $g_2$ is given by:
\begin{equation}
g_2(s)={e^{2i\omega s}\over 32\pi^2s}\big({1\over s}-i\omega\big)
\end{equation}

We consider a shell of thickness $R-r$ with inner radius $r$, and
an outer radius $R$. The permeability of the shell is $\mu_{II}$
(Fig. \ref{f2}), it is imbedded in a medium with permeability
$\mu_{III}$ and its core has permeability $\mu_I$. We define
$\kappa_r=\log{\mu_I\over\mu_{II}}$ and
$\kappa_R=\log{\mu_{II}\over\mu_{III}}$.

We now calculate the Casimir energy of the thick shell using eq.
\req{maindensity}. In our case $\mu$ is a sum of two radial step
functions, and $\mu'(s)$ is just a pair of delta functions at
$s=r$ and $s=R$. Thus the integration over $s$ in
\req{maindensity} becomes immediate and yields the energy density
\begin{equation}\label{dsrm}
{\rho_T}^{(2)}(\omega)= -4\pi\omega^2\coth({\beta\omega\over 2})
{\rm Im} \left(\kappa_r I(r)+\kappa_R I(R)\right)
\end{equation}
The $I$'s in \req{dsrm} can be explicitly calculated:
\begin{equation}
I(r)=\kappa_r\int_0^{2r}{u^3\over 2}
g_2(u)+\kappa_R\int_{R-r}^{R+r}{(u^4-(R^2-r^2)^2)\over 2u}g_2(u)
\end{equation}
and:
\begin{equation}
I(R)=\kappa_R\int_0^{2R}{u^3\over
2}g_2(u)+\kappa_r\int_{R-r}^{R+r}{(u^4-(R^2-r^2)^2)\over 2u}g_2(u)
\end{equation}
So that
\begin{eqnarray}\label{rhoTs}
& {\rho_T}^{(2)}(\omega)=-4\pi\omega^2\coth({\beta\omega\over 2})
{\rm Im} \big( \kappa_r^2\int_0^{2r}{u^3\over 2} g_2(u)+\\
\nonumber & \kappa_R^2\int_0^{2R}{u^3\over
2}g_2(u)+2\kappa_r\kappa_R\int_{R-r}^{R+r}{(u^4-(R^2-r^2)^2)\over
2u}g_2(u)\big)
\end{eqnarray}
We can identify the first two terms as the densities of Casimir
energy per $\D\omega$ for balls of radii $r$ and $R$. The total
Casimir energy can now be obtained by integration of the density
\req{rhoTs} over the frequencies $\omega$ using \req{i1} and
\req{i2}. The result is:
\begin{equation}\label{ThickShellEnergy}
E_C(r,R)=\frac{5}{128\pi R}\kappa_R^2 +
\kappa_r\kappa_R\frac{r^3(r^2-5R^2)}{4(r-R)^3(r+R)^3\pi}+\kappa_r^2\frac{5}{128\pi
r}
\end{equation}
This is our final expression for the energy of a thick shell,
where the assumption of diluteness implies
$|\kappa_r|,|\kappa_R|<<1$.

Let us check that this result coincides with the known result for
a single sphere when $R$ goes to infinity. To do so we expand
$\kappa_r$ in terms of the diluteness parameter
$\xi={\mu_I-\mu_{II}\over \mu_I+\mu_{II}}$ which is commonly used
in the literature:
\begin{equation}
\kappa_r=\log{\mu_I\over \mu_{II}}=\log{1-\xi\over
1+\xi}=-2\xi-2{\xi^3\over 3}+...
\end{equation}
Substituting this expansion in \req{ThickShellEnergy} and taking
the limit $R\rightarrow\infty$ we immediately regain the usual
result for the energy of a dielectric-diamagnetic ball
\cite{Kl,KlFe}, namely: $E_C(r,\infty)=\frac{5\,\xi^2}{32r\pi
}+{\cal O}( \xi^ 4)$. This energy yields an outward pressure on
the ball.

Note also that the Casimir energy of two parallel dielectrics per
unit area can be obtained from \req{ThickShellEnergy}, by taking
the limit of large radii. To see this we keep $d=R-r$ finite while
taking the limit $r,R\rightarrow \infty$, and divide by the
surface area thus obtaining:
\begin{equation}
E_C(d)=\frac{\kappa_r\kappa_R}{8\pi^2d^3}
\end{equation}

To illustrate the broad range of behaviors that are possible from
an expression for the energy such as \req{ThickShellEnergy}, we
study in more detail two cases:

\noindent 1. $\kappa_r=-\kappa_R$. This happens when the inner and
outer material are the same. E.g. one can imagine a thick material
shell in vacuum.

\noindent 2. $\kappa_r=\kappa_R$. In this case the ration of
magnetic constants between the inner and middle materials is the
same as the ratio between the middle and outer material.


\section{Stress}\label{STRESS}
Now we use the expression for the energy \req{ThickShellEnergy} in
order to evaluate the Casimir stress on the shells. The pressure
on the outer shell is given by
\begin{equation}
{\cal F}_R=-{1\over 4 \pi R^2}{\partial\over \partial
R}E_C={1\over 4\pi^2 R^2}\Big({r^3R(r^2+5R^2)\kappa_r
\kappa_R\over (r-R)^4(r+R)^4}+{5\kappa_R^2\over 128 R^2}\Big)
\end{equation}
While the pressure on the inner shell is given by
\begin{equation}
{\cal F}_r=-{1\over 4 \pi^2 r^2}{\partial\over \partial
r}E_C={1\over 4\pi^2 r^2}\Big({r^2(r^4-10r^2R^2-15R^4)\kappa_r
\kappa_R\over 4(r-R)^4(r+R)^4}+{5\kappa_r^2\over 128 r^2}\Big)
\end{equation}
One may recognize the self Casimir force acting on the inner and
outer shells in the terms independent of $r$ or $R$ respectively.
We now turn to examine two cases:

\subsection{$\kappa_r=-\kappa_R$}
The pressure on the outer shell is
\begin{equation}
{\cal F}_R={\kappa_R^2\over 4\pi^2 R^2}\Big({5\over 128
R^2}-{r^3R(r^2+5R^2)\over (r-R)^4(r+R)^4}\Big)
\end{equation}
The sign of this expression depends only on the ratio $c={r\over
R}$. The pressure can be written in the form:
\begin{equation}
{\cal F}_R={\kappa_R^2\over 4\pi^2 R^2}\Big({5\over 128
}-{c^3(c^2+5)\over (c-1)^4(c+1)^4}\Big)\,\,\,\,\,\,\,0<c<1
\end{equation}
\begin{figure}\center{\input epsf \epsfxsize=2.0in\epsfbox{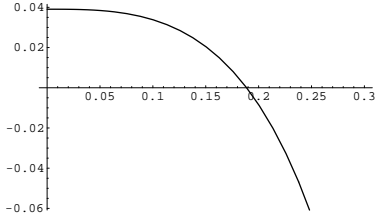}}
\caption{Pressure on outer shell as function of the radii ratio.
The Casimir force changes direction at ${r\over R}\sim 0.19$}
\label{f1}
\end{figure}
For a constant outer radius $R$, the behavior of the pressure as a
function of the ratio $c$ is indicated in Fig.\ref{f1}. One can
see that at a radii ratio of approximately 0.19 the Casimir
pressure suddenly changes sign and becomes an inward pressure.
However, the pressure on the inner shell is always directed
outward so that the inner and outer regions attract. This
attraction may be balanced by adding the compression resistance of
the middle medium, or by adding the volume dependence of $\mu$. In
the case that $r< 0.19 R$, the pressure on the inner shell is
larger than the pressure on the outer shell, so that one can
imagine the two shells growing until they reach the $0.19$ ratio
and then the outer shell will start contracting.

This result is not too surprising. If one considers two conducting
shells which are very close, then there will be attraction between
the shells, of the order of magnitude of attraction between
conducting plates. This attraction will lose its dominance once
the radii become far in magnitude and then each shell will
experience its own outward Casimir pressure. However, in order to
know if the outer radius stays finite or goes to infinity, it is
necessary to introduce the law by which the smaller radius changes
as we change the outer radius. An example for such a situation is
given in the next section. \footnote{In this section we considered
$\kappa_r=-\kappa_R$ but the qualitative results will hold
whenever $\kappa_r$ and $\kappa_R$ have different signs, as long
as they are both small and of the same order of magnitude, so as
to satisfy the validity of eq. \req{maindensity}.}
\subsection{$\kappa_r=\kappa_R$}
In this case the pressure on the outer shell is outward for all
$R$. The inner shell will be subject to an outward pressure as
long as $R>3.46r$. For $R<3.46r$ the pressure will be an inward
one on the inner shell. We return to this situation in the next
section.

\section{A thick shell with a fixed volume}\label{VOLUME}
In the previous section we calculated the pressure assuming that
the volume of the inner ball is fixed. Another interesting model
is to consider the volume of the thick shell itself as constant.
Thus we can assume $v={4\pi\over 3}(R^3-r^3)$ remains constant,
and look for a minimum of the energy under this condition.

In this case, as the outer shell expands, the distance between the
inner and outer boundaries decreases. So that if there is
attraction between the shells, it will be energetically favorable
for the shell to expand to infinity gaining from the energy of
interaction between the boundaries as well as from the tendency of
each separate boundary to grow. This is indeed what happens in the
case where $\kappa_r=-\kappa_R$.

\begin{figure}\center{\input epsf \epsfbox{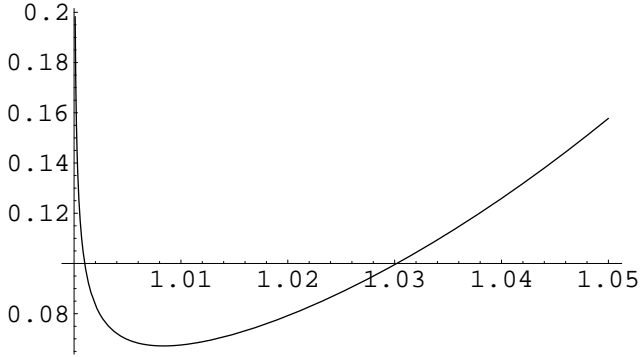}}
\caption{The potential of a thick shell with a fixed volume as
function of $R$, for $\kappa_r=\kappa_R$. There is a clear minimum
at $R=1.01$.} \label{pmin}
\end{figure}

However, for the second possibility, namely $\kappa_r=\kappa_R$
there is repulsion between the boundaries. This protects the shell
from growing to infinity and a stable minimum of the potential can
be found. A typical example is illustrated in fig. \ref{pmin},
where the minimum of the potential is for $R\simeq 1.01$, $r\simeq
0.31$ assuming the volume of the substance in the middle region is
kept constant at ${4\pi\over 3}$.

Another possibility is to keep the distance between the shells
constant (imagine the inner and outer boundaries are attached by
means of stiff rods of constant length). Qualitatively the results
are then similar to the results discussed above for the cases
$\kappa_r=-\kappa_R$ and  $\kappa_r=\kappa_R$.


\section{Discussion}
One of the most intriguing aspects of the Casimir force is its
sign: this force was shown to exhibit different behavior in
different problems. It seems that the sign of the force depends on
the balancing of several effects, such as a tendency to minimize
the average curvature of the boundaries, and on the other hand the
interaction between different patches of the boundary. In the
calculations here, which are confined to the case of uniform
velocity of light, we have shown how these effects can work with
or against each other in an explicit way resulting in a new range
of behaviors. We obtained a general expression for the energy as
function of the radii and permeabilities in the limit of dilute
media. In particular, for fixed distance between the shells, in
the limit of radii approaching to infinity, we regain, in essence,
the standard expression for the Casimir energy between parallel
dielectric media. However, now the sign of the interaction energy
in \req{ThickShellEnergy} is determined by the relative size of
their permeabilities: for a shell enclosed between two vacuua we
get attraction between the boundaries, as it should be for the
standard Casimir two plates case. Repulsion is obtained when the
permeability of the shell itself is between the permeability of
the inner substance and the permeability of the outer substance
(i.e. $\mu_I<\mu_{II}<\mu_{III}$ or
$\mu_I>\mu_{II}>\mu_{III}$)\footnote{The repulsion of plates in
such a case can be confirmed by a direct calculation in the usual
geometry of two infinite dielectric-diamagnetic media seperated by
vacuum.}.
\section{Appendix}
Two integration formulae are needed to calculate the $I$'s:
\begin{equation}\label{i1}
R_{-1}=\int{1\over u}g_2(u){\rm du}=-{e^{-2i\omega u}\over 64\pi^2
u^2}
\end{equation}
\begin{equation}\label{i2}
R_{3}=\int u^3g_2(u){\rm du}=-{e^{-2i\omega u}\over 32\pi^2
}({u^2\over 2}-{1\over 2\omega^2}-i{u\over \omega})
\end{equation}
\section*{Acknowledgments}
I.K. wishes to thank Joshua Feinberg and Oded Kenneth for
discussions. This work was supported by the fund for promotion of
research at the Technion and by the Technion VPR fund.
 
\end{document}